\documentstyle[galley,epsfig]{mn2e}
\title{HD~209621: Abundances of neutron-capture elements}
\title[HD~209621: Abundances of neutron-capture elements]
{HD~209621: Abundances of neutron-capture elements\thanks
{Based  on data collected at the Subaru Telescope, which is 
operated by the National Astronomical Observatory of Japan  and 
at HCT, IAO, Hanle, India}}

\author[Aruna Goswami et al. ]{Aruna Goswami$^{1}$, Wako Aoki$^{2}$  \\
    $^{1}$Indian Institute of Astrophysics, Koramangala, Bangalore 560034,
    India; aruna@iiap.res.in \\
    $^{2}$National Astronomical Observatory, Mitaka, Tokyo, 181-8588 Japan;
    aoki.wako@nao.ac.jp\\
}    
\begin{document}

\date{ Accepted 2009 December 30;  Received 2009 December 24;  in original 
form 2009 October 28 \large \bf }

\pagerange{\pageref{firstpage}--\pageref{lastpage}} \pubyear{2009}

\maketitle

\label{firstpage}

\begin{abstract}
High resolution spectra obtained from the Subaru Telescope High Dispersion 
Spectrograph have been used to update the stellar atmospheric parameters 
and metallicity of the star HD~209621. We have derived a metallicity 
of ${\rm [Fe/H]} = -1.93$  for  this star, and have found a large 
enhancement of carbon and of heavy elements, with respect to iron. Updates 
on  the elemental abundances of four s-process elements (Y, Ce, Pr, Nd)
along with the  first estimates of abundances for a number of other 
heavy elements (Sr, Zr, Ba, La, Sm, Eu, Er, Pb) are  reported.  The 
stellar atmospheric parameters, the effective temperature, T$_{eff}$, 
and the surface gravity, log\,$g$ (4500 K, 2.0), are determined from 
LTE analysis using  model atmospheres. Estimated [Ba/Eu] = +0.35, 
places the star in the group of CEMP-(r+s)  stars; however, the 
s-elements abundance pattern seen in HD~209621 is  characteristic of 
CH stars; notably, the 2nd-peak s-process elements are more enhanced 
than the first peak s-process elements. HD~209621 is also found 
to show a large enhancement of the 3rd-peak s-process element lead (Pb) 
with ${\rm [Pb/Fe]} = +1.88$. The relative contributions of the two 
neutron-capture processes, r- and s- to the observed abundances are examined 
using a parametric model based analysis, that hints that the neutron-capture 
elements in HD~209621 primarily originate in s-process.
\end{abstract}


\section{Introduction}

The population II CH stars with their characteristic properties like iron 
deficiency and enhancement of  carbon and s-process elements can provide strong 
observational constraints for the theoretical computation of  
nucleosynthesis at low metallicity. Early-type extrinsic  CH stars
 ($^{12}$C/$^{13}$C ${\le}$10)  are confirmed post-mass-transfer 
binaries (McClure \& Woodsworth 1990) in which the companion star is 
an AGB that evolved to a now invisible white dwarf. The chemical 
composition of the early type CH  stars  that are characterized by 
enhancement of s-process elemental abundances bears signatures of the 
nucleosynthesis processes operating in low-metallicity  companion AGB 
stars, provided they conserve the surface  characteristics of the 
companion  stars. These stars thus form ideal  targets for studying 
the operation of  s-process at low metallicity. However, not many studies  
on CH stars can be found; the few earlier studies available are either 
limited by the resolution or by the  wavelength range. It is important to 
update the elemental abundances based on higher resolution spectra; the 
results obtained for heavy elements Zr, La, Ce, Nd with better spectra 
in a number of CH stars are found to be very different (van Eck et al. 
2003) from those estimated earlier (Vanture 1992c). In the present work 
we report  new results on the chemical composition of HD~209621
 (V$^{*}$HP Peg)  listed  in the CH star catalogue of 
Bartkevicius (1996).    

The star HD 209621 along with two other CH stars HD~5223 and HD~26  have
 been used as   reference  stars
 for studies on medium resolution spectroscopic analysis of candidate
Faint High Latitude Carbon stars from the Hamburg/ESO survey
(Goswami 2005, Goswami et al. 2007, Goswami et al. 2009). 
While fairly detailed abundance analyses are available on the stars 
HD~26 (Van Eck et al. 2003) and HD~5223 (Goswami et al. 2006) a detailed 
chemical composition for HD~209621 is lacking. Earlier studies
 on this object are  limited by  both resolution as well as 
wavelength regions.  Thus the nature of the distribution of the
neutron-capture elements and their abundances for this star are
necessary to justify its further use as a reference CH star; the present 
work accomplished this using a high resolution Subaru spectrum for this star.

Estimated $^{12}$C/$^{13}$C $\sim$ 10 (Tsuji et al. 1991)  places HD~209621  in
the group of early-type CH stars. A carbon isotopic ratio  of 4 is reported
 by Climenhaga (1960); Vanture (1992a) estimated this value  to be $\sim$ 3 
using three spectral features: the $^{12}$C$^{13}$C isotopic 1-0 Swan bandhead
at 4744 {\AA}, a blend of three  $^{13}$CN lines at  8004 {\AA}, and a 
 $^{13}$CN feature at 8036 {\AA} which is also a blend of three $^{13}$CN 
lines.  Goswami (2005) reported a carbon isotopic ratio of 8.8 measured on 
a medium resolution spectrum of this object using molecular band-depths of
(1,0) $^{12}$C$^{12}$C ${\lambda}$4737 and
 (1,0) $^{12}$C$^{13}$C ${\lambda}$4744.
  
 From a comparison of  the star's spectrum (13.5 \AA\, mm$^{-1}$ 
spectrograms) with that of  ${\epsilon}$Vir, Wallerstein (1969) reported an 
effective temperature of 4700 K for this object.  The  comparison star 
${\epsilon}$Vir, is  a  high proper motion star of spectral type G8 III,
with near solar metallicity [Fe/H]=0.15 (McWilliam 1990).  With respect to
 ${\epsilon}$Vir,  Wallerstein derived   a metal-deficiency for HD~209621 by 
a factor of twenty and an  enhancement of the ratios of rare earths to metals
 by a factor of eight. Despite  the low metal abundance and  high space 
velocity (velocity components U, V, W = -69, -332, +176), the  star was
 found to lie  to the right and below  the vertical or giant branches of 
the globular clusters (Wallerstein 1969).

 Vanture (1992c), adopting  an effective temperature of 4700 K from Wallerstin
 (1969)  determined abundances for four  s-process  elements Y, Ce, Pr and Nd
  using its spectra at resolution  ${\lambda}/{\delta\lambda}$ = 20\,000, 
lower than  ours (${\lambda}/{\delta\lambda}$ = 50\,000). Another objective of
 the present work is to examine and update 
 the  elemental  abundances  of these four heavy elements 
  and estimate  abundances for     other heavy 
elements like Sr, Zr, Ba, La, Sm, Eu, Er, W and  Pb. 

 Details of observation  and  data reduction  are reported in section 2. In  
section 3, we  present $BVRIJHK$ photometry  and discuss the  estimates of 
the effective temperatures  from photometry. Determination of the radial 
velocity and the stellar atmospheric parameters are discussed in 
section 4.  The elemental abundance results are presented in section 5.
A brief discussion on the parametric model analysis of the observed abundances
is presented   in section 6. Discussion and concluding remarks are drawn
 in section 7. 

\section{Observation and  Data Reduction}

High-resolution spectroscopic observations of HD~209621 was carried out with 
the High Dispersion Spectrograph (HDS) of the 8.2m Subaru Telescope (Noguchi 
et al. 2002) on 20 July, 2005  at a  resolving power of $R \sim$ 50\,000. The 
object spectrum was taken with a 5  minutes exposure.  The observed bandpass 
ran from about $4020$\,{\AA} to $6775$\,{\AA}, with a gap of about 
$75$\,{\AA}, from $5335$\,{\AA} to $5410$\,{\AA}, due to the physical spacing 
of the CCD detectors.  Observations of a Th-Ar hollow cathode lamp,
 provided the
wavelength calibration. Standard spectroscopic reductions (e.g., flat
fields, bias subtraction, extraction, and wavelength calibration) were
carried out using the IRAF\footnote{IRAF is distributed by the National
Optical Astronomical Observatories, which is operated by the Association
for Universities for Research in Astronomy, Inc., under contract to the
National Science Foundation} spectroscopic reduction package.

\section{Photometry and Effective temperatures}

Optical broadband $BVRI$ photometry along with near-IR $JHK$ photometry from 
2MASS (Skrutskie et al. 2006), are listed in Table 3.  The reddening estimate 
for HD~209621, $E(B-V) = 0.09$,  is  from  Sleivyte and Bartkevicius (1990). 
$B-V$, $U-B$, $R-I$ and $V-I$  values are taken from Platais et al. (2003).

We have used colour-temperature calibrations of Alonso et al. 1996) that 
relate $T_{\rm eff}$ with various optical and near-IR colours. Estimated 
uncertainty in the temperature calibrations  is $\sim 90$\,K. The Alonso et 
al. calibrations use Johnson photometric systems for UBVRI and  use TCS 
(Telescopio Carlos Sanchez) system  for IR colours, J-H and J-K. 
The necessary transformations 
between these photometric systems are performed using transformation relations 
from Carpenter (2001), Bessell and Brett (1988) and Alonso et al. (1996, 1999). 
The colour-based estimates of $T_{\rm eff}$ agree well within 100 - 200 K 
except for the temperature estimated from $B-V$ colour.  The $B-V$ colour of a 
star with strong molecular carbon absorption features depends not only 
on $T_{\rm eff}$, but also on the metallicity of the star and on the strength 
of its molecular carbon absorption features, due to the effect of CH molecular 
absorption in the B band. We have not  used the empirical $T_{\rm eff}$ scale 
for the $B-V$ colour indices. The derived $T_{\rm eff}$ from V-K is ${\sim}$ 
400 K, and from J-H  is ${\sim}$ 250 K less than the adopted spectroscopic
 $T_{\rm eff}$   derived by imposing Fe I excitation equilibrium, described 
in the next section. The temperature calibrations  from the
 $T_{\rm eff}$ - $(J-H)$ and $T_{\rm eff}$ - $(V-K)$ relations  involve a 
metallicity ([Fe/H]) term. Estimates of  $T_{\rm eff}$  
at  two assumed  metallicity values (shown in paranthesis)  are listed in 
Table 4.
\begin{figure*}
\epsfxsize=10truecm
 \epsffile{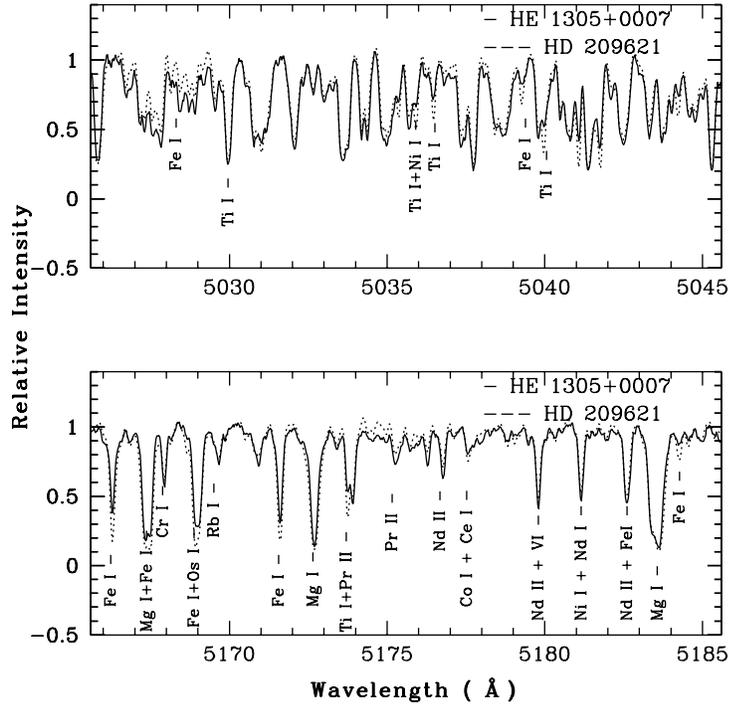}
\caption{A comparison of the spectra  of HD~209621 and  HE~1305+0007
 in the wavelength region $5025.6$\,{\AA} to $5045.6$\,{\AA} (upper panel)
and  $5165.6$\,{\AA} to $5185.6$\,{\AA} (lower panel).
A number of prominent lines are indicated on the spectra; the features in 
HD~209621 closely match their counterparts in HE~1305+0007.  }
\end{figure*}

\begin{figure*}
\epsfxsize= 9truecm
 \epsfig{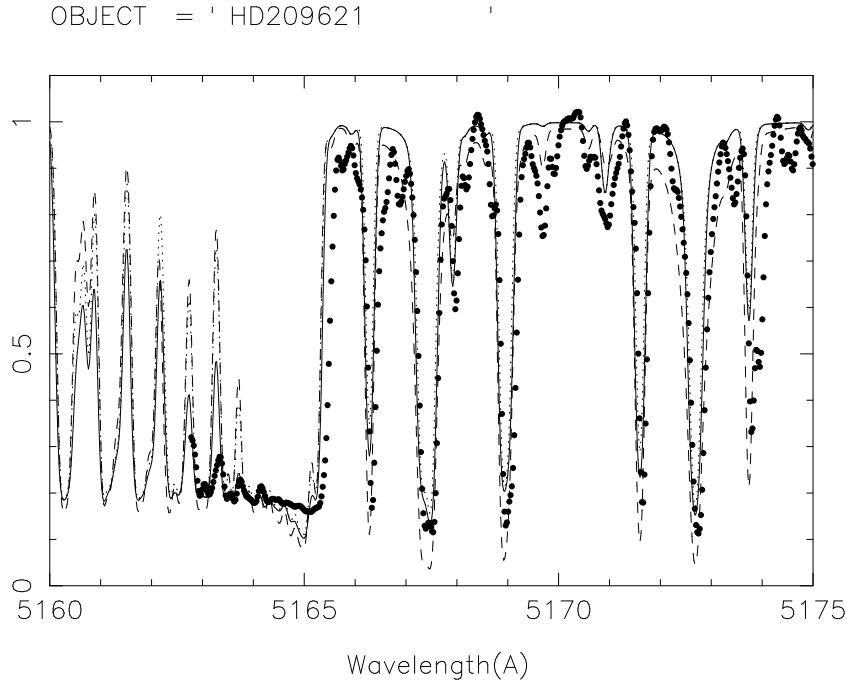}
\caption{ A fit of the synthetic spectrum (dotted curve) compared with the
observed spectrum (solid curve) of HD~209621 
in the wavelength region 5160 to $5175$\,{\AA}. The synthetic
spectrum is obtained using a model atmosphere corresponding to the adopted
parameters listed in Table 5. Calculations  made
using Vanture's parameters (dashed curve)
 result in too strong absorptions (in particular Mg line wings).}
\end{figure*}

\section{Description of the spectra}

The high-resolution spectra of HD~209621 are
characterized by  closely-spaced molecular absorption lines of CH, CN
and the Swan system of C$_{2}$. The continuum is obscured over essentially
the entire wavelength region.
The spectra of HD~209621 are dominated by the lines of carbon bearing
molecules;  a few unblended atomic lines were possible to identify only from
the regions that are relatively clear of molecular lines. 
Possible blends between atomic and molecular lines were identified on a 
line-by-line basis using the solar atlas, CH lines and C$_{2}$ Swan band
systems line lists of Phillips and Davis (1968) and in consultation
with atomic line lists of Kurutcz. All lines suspected of being blends
were eliminated;
 a final list of Fe I and Fe II lines 
considered in the present analysis  for the determination of atmospheric
parameters and metallicity  is given in Table  1.  
Two spectral regions  of the star 5025-5045 and  5165-5185 \AA\,  are shown in 
 Figure 1  to illustrate the complexity
of the star's spectra. A comparison of the spectral regions  with those of 
HE~1305+0007, a double enhanced star with comparable metallicity
  ([Fe/H] = -2.05, Goswami et al. 2006),  shows great similarity.

{\footnotesize
\begin{table*}

{\bf Table 1:  The list of Fe lines used in the present analysis  }\\

\begin{tabular}{ c c c c c   c }
         &    &           &          &           \\
\hline
         &    &           &          & HD~209621 \\
W$_{lab}$& ID & EP$_{low}$& log${gf}$& Eq widths (m\AA\,) \\
  
\hline
 6230.726&Fe I&   2.5590  &  $-$1.281  &  141.3    \\ 
 6137.694&Fe I&   2.5881  &  $-$1.403  &  127.4     \\ 
 6136.615&Fe I&   2.4530  &  $-$1.400  &  147.2    \\ 
 5586.760&Fe I&   3.3683  &  $-$0.210  &  125.7    \\
 5324.178&Fe I&   3.211   &  $-$0.240  &  139.0    \\
 5324.179&Fe I&   3.211   &  $-$0.103  &  139.0    \\
 5266.555&Fe I&   2.9980  &  $-$0.492  &  138.7   \\
 5242.490&Fe I&   3.6300  &  $-$0.840  &   64.6    \\
 5232.939&Fe I&   2.9400  &  $-$0.190  &  154.6   \\
 5202.340&Fe I&   2.1800  &  $-$1.840  &  134.0    \\
 5192.343&Fe I&   2.9980  &  $-$0.521  &  124.0    \\
 5171.595&Fe I&   1.4848  &  $-$1.793  &  174.7     \\
 5006.119&Fe I&   2.8330  &  $-$0.610  &  134.6   \\
 5001.860&Fe I&   3.8800  &  $+$0.009  &   80.9   \\
 4924.770&Fe I&   2.2786  &  $-$2.222  &   94.5    \\
 4918.990&Fe I&   2.8700  &  $-$0.340  &  139.9    \\
 4871.317&Fe I&   2.8650  &  $-$0.410  &  142.8    \\
 4484.219&Fe I&   3.6025  &  $-$0.720  &   66.5    \\
 5234.625&Fe II&  3.2214  &  $-$2.050  &   74.8    \\
 4923.930&Fe II&  2.8910  &  $-$1.319  &  144.3    \\
 4583.839&Fe II&  2.8070  &  $-$2.020  &   93.8    \\
 4508.280&Fe II&  2.8555  &  $-$2.210  &   60.1   \\
\hline  

\end{tabular}
\end{table*}
}

\subsection{Radial velocity}

We have estimated the radial velocity of   HD~209621 using several 
unblended lines.  Estimated heliocentric radial velocity $v_{\rm r}$ 
is listed  in Table 2.
 The SIMBAD database lists a mean radial velocity of 
 $v_{\rm r} =-381$ km s$^{-1}$ for HD~209621.  The star is a radial velocity
variable. McClure and Woodsworth (1990)
determined a period of 407.4 days for this star.

{\footnotesize
\begin{table*}

{\bf Table 2:  Heliocentric Radial velicity    v$_{r}$  of HD~209621}\\

\begin{tabular}{ l c c c c  }
  &   &   &  &    \\

\hline

Star Name  & $v_{\rm r}$ km s$^{-1}$  & HJD    &  $v_{\rm r}$ km s$^{-1}$ \\
        & our estimation     &        & from literature    \\ 
\hline
HD~209621      &$-$390.5 ${\pm}$ 1.5 & 2453541.95451 & $-$381     \\
\hline
\end{tabular}
\end{table*}
}

{\footnotesize
\begin{table*}

{\bf Table 3: Photometric parameters of HD~209621}\\

\begin{tabular}{ l c c c c c c c c c c  c c}
  &   &   &  &  &   &  &  &  &   &  \\

\hline

Star Name   &RA(2000)  & Dec(2000)     &  V   & $B-V$   &$U-B$ & $R-I$  & $V-I$   &$E(B-V)$
&  $J$    &$ H$  & $K_{s}$\\
\hline
HD~209621 & 22 04 25.14 & $+$21 03 09.0 & 8.86 & 1.45& 1.14&  0.61  & 1.15  & 0.09  &6.661   & 6.045  & 5.913 \\

\hline

\end{tabular}
\end{table*}
}

{\footnotesize
\begin{table*}
{\bf Table 4: Estimated effective temperatures ($T_{\rm eff}$) from 
semi-empirical relations }\\
\begin{tabular}{lcccc}
\hline
Star Name  &$T_{\rm eff}$ &$T_{\rm eff}$ &$T_{\rm eff}$ &   \\
            &  $(J-K)$        &  $(J-H)$    &  $(V-K)$    &        \\
\hline
HD~209621    &  4205.8   & 4190.4 ($-$1.0) & 4193.4 ($-$1.0)  &         \\
            &           & 4223.8 ($-$2.0) & 4102.1 ($-$2.0)  &            \\
\hline

\end{tabular}
 \\
The numbers inside the parentheses indicate the adopted metallicities [Fe/H] \\
\end{table*}
}

\subsection{Stellar atmospheric parameters }
The stellar atmospheric parameters, the effective temperature 
($T_{\rm eff}$), the surface gravity (log$g$), and metallicity 
([Fe/H]) of the star are  determined by an LTE analysis of equivalent
widths of  atomic  Fe lines using a  recent version of MOOG of Sneden 
(1973). Eighteen cleanest possible lines of  Fe I  and  four Fe~II 
lines (Table 1)  are used in  our analysis.  Model atmospheres are 
selected from the Kurucz grid of model atmospheres computed with 
better opacities and abundances  with no convective overshooting. 
 These models are available at {\tt http://cfaku5.cfa.harvard.edu/}, 
labelled with the suffix ``odfnew''. The excitation potentials and
oscillator strengths of the lines are  taken from various sources,
including the Vienna Atomic Line 
Database ({\tt http://ams.astro.univie.ac.at/vald/}),  Kurucz atomic 
line list ({\tt
http://www.cfa.harvard.edu/\-amp/\-ampdata/\-kurucz23/\-sekur.html}),
Fuhr, Martin, \& Wiese (1988), Martin, Fuhr, \& Wiese (1988), and Lambert
et al. (1996). The $gf$ values of  elements compiled by  
 R.E.  Luck (private communication) are also consulted.

The effective temperature have also been  obtained by the method of
excitation balance, forcing the slope of the abundances from Fe~I lines
versus excitation potential to be near zero.  The photometric parameters
of HD~209621 are presented in Table 3. The temperature estimates 
derived from JHK photometry (Table 4)  provided a preliminary temperature 
check corresponding to which initial model atmosphere is  chosen.
The  effective temperature is then obtained by an iterative
process using the method of excitation balance.   Estimated
$T_{\rm eff}$ is 4500 K.  Tsuji et al (1991)  derived $T_{\rm eff}$ $\sim$ 
4400 K  for this object based on Infra Red  Flux Method (IRFM), that 
falls well within the error limits of our temperature determination.
We have adopted a microturbulence of 2 km s$^{-1}$ for this star.
 Such a value  is not unrealistic; (in cool giants, with log\,$g \le$ 2.0, 
in general $V_{\rm t}$  ${\ge}$ 2 km s$^{-1}$  (Vanture 1992c, 
McWilliam et al. 1995a,b)). Using the Fe~I/Fe~II ionisation equilibrium,
 the surface gravity of HD~209621 is obtained as  log $g$ = 2.0, same as 
the value adopted by Tsuji et al. (1991). The metallicity of the star 
is estimated as [Fe/H] = -1.93, significantly lower than the metallicity
derived by Vanture (1992b,  [Fe/H] = -0.9).

In figure 2, we have shown a fit of the  synthetic spectrum (dotted curve) 
compared with the observed spectrum (solid curve) of HD~209621
in the wavelength region $5163$ to $5175$\,{\AA}. The region is particularly
chosen being  relatively free from contamination by molecular features. 
The synthetic spectrum is obtained using a model atmosphere corresponding 
to the adopted parameters listed in Table 5. Calculations are also made 
using the set of   parameters (v$_{t}$ = 3.25 solid line), changed 
microturbulence; (v$_{t}$ = 2.0 dotted line)  and for Vanture's parameters;
 Vanture's parameters result in too strong absorption (in particular Mg 
line wings). This indicates that Vanture (1992b) had estimated a higher 
metallicity for this star. The  metallicity derived by us ([Fe/H] ${\sim}$ $-$1.93)
is more likely for HD~209621. The atmospheric parameters, the effective 
temperature and the surface gravity determined by him, however, do not 
differ much from the ones derived by us.  Using  $T_{\rm eff}$ = 4700 K of 
Wallerstein (1969), we could not achive a  satisfactory fit, it seemed 
necessary to adopt instead a somewhat lower value for $T_{\rm eff}$. 

From an arithmatic mean of two estimates one (`bolometric') from a 
reference integrated flux F$_{0}$, the other (`spectral') from calibrated
color indices which are representative of Spectral Energy Distributions (SED)
shapes, Bergeat et al. (2001) proposed an effective temperature of
4190 K for HD~209621. This value also did not give a satisfactory fit.

The correctness of our estimates is varified by reproducing the atmospheric 
parameters obtained by Barbuy et al. (2005)  for the star CS~22948-027.
Their  estimates are  very close to our  estimates  (refer to Goswami et al.
 2006, Table 5).  In our  analysis that follows,  we have adopted 
  $T_{\rm eff}$  = 4500 K,  log\,$g$ = 2.0,
 and  $V_{\rm t}$ = 2.0 km s$^{-1}$.

\section{Abundance Analysis}

Due to severe line blending  throughout the spectral range 
a standard abundance analysis procedure based on the equivalent widths 
could not be applied  for elements other than iron
Therefore, the  elemental abundances  are  derived  from spectrum-synthesis 
calculations. The same  method  is also applied for all lines irrespective
of whether they are affected by  hyperfine splitting or not.
Local thermodynamic equilibrium is assumed for  the
spectrum-synthesis calculations. We have used the latest version of MOOG
Sneden (1973)
for spectrum synthesis.  The line list for each region  synthesized
is taken from the Kurucz atomic line list ({\tt
http://www.cfa.harvard.edu/\-amp/\-ampdata/\-kurucz23/\-sekur.html}) and
from the
Vienna Atomic Line Database ({\tt http://ams.astro.univie.ac.at/vald/}).
Reference solar abundances for the various elemental species  are adopted from
Asplund, Grevesse \& Sauval (2005). The log\,$gf$ values for atomic lines are
also adopted from Fuhr et al. (1988) and Martin et al. (1988) whenever 
available,
and  from a compilation of $gf$ values by R. E. Luck 
(private communication). For heavy neutron-capture elements
log\,$gf$ values 
given by Sneden et al. (1996) and Lawler et al. (2001) are  also consulted.
The lines used for  abundance determination  of the elements  and their
atomic parameters are listed in Table 6. An extended line-list of  248 lines
 of heavy elements is  available (Table 9)  as on-line material.  
Information on the 
 atomic line profiles extracted from the spectrum of HD 209621 has been 
listed in that table. Table 6
      lists only those lines that have been used for abundance determination.
The derived abundances are presented in  Table 7.
 In this
Table we have listed  the abundance log $\epsilon$(X), along with [X/H]
and  [X/Fe] values. In computing the quantity [X/Fe]  we have used the
Fe~I-based abundance for elemental abundances derived from neutral lines
and the Fe~II-based abundance for elemental abundances derived from ionized
lines.  
Estimated elemental abundances  listed in Table 7  are discussed below.  

\subsection{ Carbon, Nitrogen, Oxygen}

{\it Carbon} (C)\,  Carbon abundance is derived from spectral 
synthesis of the C$_{2}$
Swan 0-1 band around $5635$\,{\AA}. A synthetic spectrum, derived with
an  appropriate model atmosphere  and
using a
carbon abundance of log ${\epsilon}$(C) = 7.7 ${\pm}$ 0.2, 
 shows a good match to the depth of the
observed spectrum of HD~209621. Relative to the solar photospheric C
abundance, C is strongly enhanced in HD~209621 (${\rm [C/Fe]} = +1.25$ ). 

The  G band of CH is severely saturated  in the spectra of HD~209621. This
feature  is found
to be quite insensitive to the carbon abundance. Spectrum synthesis of this
feature is likely to return uncertain values and hence  we have  not considered
 this band for deriving C abundance.
Spectrum synthesis  fits of the  C$_{2}$ Swan 0-1 band around
$5635$\,{\AA}\ is found to provide  reasonable estimate.

Reported C, N, O abundances in HD~209621 by Vanture (1992c) are respectively
8.5, 8.2 and 8.4.  Oxygen abundance  by Vanture (1992c) is determined from 
OI near-infrared triplet near  7774 \AA\, applying non-LTE corrections.
No clean oxygen lines are  detected in our spectra. We have made a rough 
estimate of
oxygen by considering the fact that, the CH stars for which oxygen 
abundances have been determined  follow a particular trend:  Oxygen
abundance in halo stars are found to increase with decreasing metallicity
as [O/Fe] ${\sim}$ $-0.5$ [Fe/H] until [Fe/H] = -1.0 and then level off 
at a value of [O/Fe] = +0.35 ( Wheeler et al. 1989). 
Based on this we  have noted an Oxygen abundance of 7.15
(log$\epsilon$(O) = 7.15). 
Also, by assuming the general trend of  [C+N/Fe] = 1  observed in other  CH 
stars  and by using our estimated carbon abundance  we have derived the 
nitrogen  abundance as log$\epsilon$(N) = 8.4.
  This shows a nitrogen enhancement of  [N/Fe] = +2.5. Vanture estimated 
log$\epsilon$(N) = 8.2, indicating  [N/Fe] = +2.2.

\subsection{The odd-Z elements Na and Al }

{\it Sodium} (Na)  and {\it Aluminium} (Al) are two monoisotopic, 
odd elements.  Na abundance in HD~209621 is calculated from the
resonance doublet --  Na I D lines at $5890$\,{\AA} and
$5896$\,{\AA}. These resonance lines are sensitive to non-LTE effects
(Baum\"uller \& Gehren 1997; Baum\"uller et al. 1998; Cayrel et
al. 2004). The derived abundance from an LTE analysis is near solar
 ${\rm [Na/Fe]} = +0.01$.

Observations of  Stephens (1999) suggest,  Na/Fe decreases with decreasing 
 [Fe/H]  from  -1 to -2, as expected theoretically. Most other observations 
available in literature do not support this trend showing instead a flat 
[Na/Fe] ${\sim}$ 0 ratio with a large scatter. 

Al  line at $3961.5$\,{\AA} which  is  generally used to derive Al abundances
 in extremely metal-poor stars is out of the spectral coverage of HD~209621.
Other Al lines  towards red  region are severely blended 
 and could not be used for abundance determination. 
Although common in Globular clusters
enhancement of Na and Al is  commonly observed  in Globular clusters, but
in general, not  in field metal-poor stars.

\subsection{The ${\alpha}$-elements  Mg, Si, Ca, and Ti }

{\it Magnesium} (Mg)\, The Mg abundance is derived from the synthesis 
of the
Mg~I line at 5172.68 {\AA}. The predicted line profile with our adopted Mg
abundance (log $\epsilon$(Mg) = 5.76) fits the observed line profile in
HD~209621 quite well. Magnesium is found to exhibit an overabundance  with
 [Mg/Fe] $\sim +0.17$. 

{\it Silicon} (Si)\, Lines of Si detected in the spectra of HD~209621 are
severely blended and could not be used for abundance determination. 

{\it Calcium} (Ca)\,   The  abundance of Ca is derived from
 the Ca~I line at $6102.7$\,{\AA}. In HD~209621, Calcium 
abundance gives   ${\rm [Ca/Fe]}$ = +0.08.
The Ca abundance  is practically as expected for a halo star with
${\rm [Fe/H]}$ = -2.0 (Goswami \& Prantzos 2000, Figure 7).  

{\it Titanium} (Ti) The Ti  abundance
derived from the synthesis of the Ti~II line at $4805.085$\,{\AA} shows a
marked overabundance of  +0.72 with respect to Fe.

\subsection{The iron-peak elements Cr, Mn, Ni, Zn}

{\it Cromium} (Cr)\,   A spectrum synthesis fit  of the
Cr~I line at $5247.566$\,{\AA}  returned an abundance estimate of 
  log${\epsilon}$(Cr) = 3.5.  Cr is mildly deficient  with 
${\rm [Cr/Fe]}$ = $-$0.20 .  Cr I line at 4254.332 \AA\, 
with a zero lower excitation potential ( E$_{low}$)  is blended with
 a Ca I line. Lines at 4591.389,
4626.174, 5206.038, 5345.801 \AA\,  with similar E$_{low}$ as that of 
$5247.566$\,{\AA}  appear  blended.

{\it Manganese} (Mn)\, No clear good lines due to Mn  are  detected in 
the spectra. Mn I lines at 5516.774 \AA\,, 
 4766.41  \AA\, and 4727.46 \AA\,  are detected as blended lines.

{\it Nickel} (Ni)\,  The  abundance  of Ni is  derived from a spectrum synthesis
  calculation  of the
Ni~I line at $5081.107$\,{\AA}. The measured equivalent width of this line 
is  37 m\AA\,.  
 Ni~I lines  at 5578.711 \AA\,  and 5754.655 \AA\,
are  detected as blended lines.  Nickel exhibits a near solar abundance of
${\rm [Ni/Fe]}$ = -0.01.

{\it Zinc} (Zn)\, The Zn abundance is derived from  a spectrum synthesis 
calculation of 
Zn I line at 4810.528 \AA\,. Zn is found to be  mildly enhanced with
[Zn/Fe] = 0.24.

\subsection{The light s-process elements Sr, Y, and Zr}

{\it Strontium} (Sr)\, The Sr abundance is derived from Sr I line  
at  $4607.327$~{\AA}.
Sr shows an overabundance of ${\rm [Sr/Fe]} = +1.02$;
log $gf$  value (-0.57) for this line is taken from Corliss and Bozman (1962).
If we adopt a log$gf$ value of +0.28 (Sneden et al. 1996) for
this line  we derive a near-solar abundance 
for Sr with [Sr/Fe] = +0.08. 

{\it Yttrium} (Y)\, The abundance of Y is derived from  $5200.41$\,{\AA} line;
Y shows an overabundance of ${\rm [Y/Fe]} = +0.36$.
 Our  adopted log$gf$ value for this line is
same as used by Sneden et al. (1996). 

{\it Zirconium} (Zr)\, The  abundance of Zr is derived using Zr I line
at $6134.57$\,{\AA}.  Zr shows an overabundance of  ${\rm [Zr/Fe]} = +1.80$.
log $gf$ value for this line is taken from Biemont et al. (1981).

Spectrum synthesis fits of Sr, Y and Zr are shown in Figure 3. A discussion 
on the line selection  of the light s-process elements are provided in APPENDIX - A. \\

\subsection{The heavy $n$-capture elements:  Ba, La, Ce, Pr, Nd, 
Sm, Eu, Er, W, Pb}

The abundances of the heavy $n-$capture elements  barium (Ba),
lanthanum (La), cerium (Ce), neodymium (Nd),  samarium (Sm), 
praeseodymium (Pr),
europium (Eu), erbium (Er), tungsten (W) and lead (Pb) are 
determined using spectrum-synthesis calculations. Spectrum synthesis fits
 for Ba, La, Nd and Sm are shown in figure 4.  The derived abundances 
are generally found to be overabundant with respect to Fe.
 Hyperfine-splitting corrections 
for  Y, Ce, and Nd lines are not used.  The abundances are 
not expected to be affected by hyperfine splitting  when derived from
 weak lines (McWilliam et al. 1995a, b).  Line
selection of the heavy $n$-capture elements for abundance determination 
is discussed in APPENDIX - A. 

{\it Barium} (Ba)\,  The  Ba II line at 6141.727 \AA\, is used 
 to determine the Ba abundance. Although broad, this
line appears as a well defined symmetric line compared to  the other barium
lines detected in the spectrum. log$gf$ value for this line is taken from
Miles and Wiese (1969). 

{\it Lanthanum} (La)\,  The abundance of La is derived from a 
spectrum-synthesis calculation
of the La~II line at $5259.38$\,{\AA}, with atomic data taken from Lawler
et al.  (2001).  La  exhibits an overabundance of
 ${\rm [La/Fe]} = +2.41$.

{\it Cerium} (Ce)\, 
The abundance of  Ce is derived from
a spectrum-synthesis calculation of the Ce~II line at $5274.23$\,{\AA}. 
Ce exhibits a large
 overabundance of  ${\rm [Ce/Fe]} = +2.04$; this value is  lower than
 (+2.8 $\pm$ 0.3) derived for Ce by Vanture (1992c).

{\it Praeseodymium}  (Pr)\,  The abundance of Pr is derived using
 Pr~II line at $5259.7$\,{\AA}; Pr too
exhibits a large overabundance of ${\rm [Pr/Fe]} = +2.16$. This estimate is
similar to the value of  +2.2 $\pm$ 0.7 derived for Pr by  Vanture (1992c).

{\it  Neodymium } (Nd)\, The abundance  of Nd is  derived using the 
Nd~II line at
$5825.85$\,{\AA};  Neodymium  shows  overabundance 
 with ${\rm [Nd/Fe]} = +1.87$; this value is also  lower than  Vanture's 
estimate of +2.4 $\pm$ 0.6. 

{\it Samarium }  (Sm)\, The abundance of Sm is derived from the  Sm~II line at
$4791.60$\,{\AA}.   The log$gf$ value is taken from Lawler et al. (2006).
 Sm exhibits an overabundance of ${\rm [Sm/Fe]} = +1.84$. 
 
{\it Europium } (Eu)\, 
The abundance of
Eu  is  determined from the red line at $6437.64$\,{\AA}.  Eu is 
overabundant with ${\rm [Eu/Fe]} = +1.35$. 

{\it Erbium } (Er)~ The  abundance of Er  is estimated using
Er II line at  4759.671 {\AA}\,. The line parameters adopted from 
Kurucz atomic line list come from Meggers et al. (1975). Er shows an 
overabundance with [Er/Fe] = +2.06.

{\it Tungsten } (W)~ The abundance of W is determined using the W I line
at 4757.542 {\AA}\,. The line parameters  taken from the  Kurucz  
atomic line list  come from Obbarius and Kock (1982). The abundance derived
is high  with [W/Fe] = 2.61; there is a possible  blend with Cr I line at
4757.578 \AA\,  (with log $g$ and lower excitation potential, -0.920 and 3.55 
respectively) resulting  in an  overestimate of  W abundance.
 
 {\it Lead } (Pb)~  Spectrum-synthesis calculation is also used
to determine the abundance of Pb  using the Pb I line at
$4057.8$\,{\AA}.  Pb shows an overabundance with
 ${\rm [Pb/Fe]} = +1.88$.

\begin{figure*}
\epsfxsize=  9truecm
 \epsffile{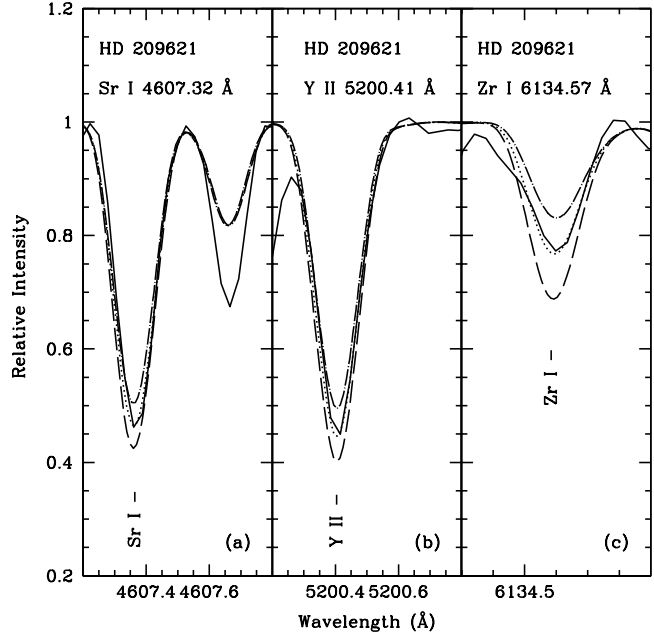}
\caption{Spectral-synthesis fits of absorption lines arising from the light
s-process elements Sr, Y, and Zr, obtained with the elemental abundances
listed in Table 7.  The dotted lines indicate the synthesized spectra and
the solid lines indicate the observed line profiles. Two alternative
synthetic spectra for ${\Delta}$[X/Fe] = +0.3 (long dash) and
${\Delta}$[X/Fe] = $-$0.3 (dot-dash) are  shown to demonstrate the
sensitivity of the line strength to the abundances. }
\label{Figure 3 }
\end{figure*}

\begin{figure*}
\epsfxsize= 9truecm
 \epsffile{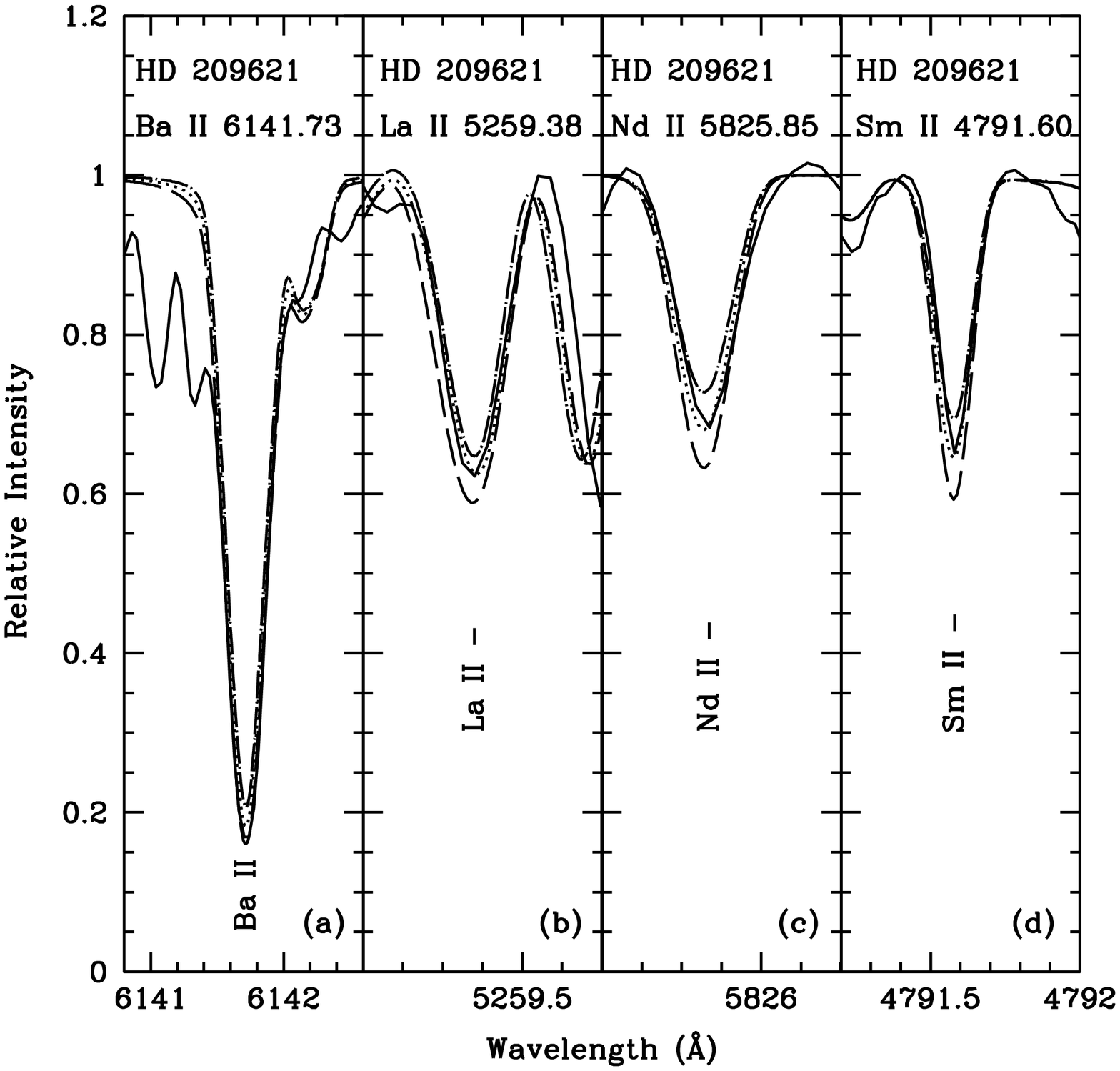}
\caption{Spectral-synthesis fits of absorption lines arising from the
heavy s-process elements Ba, La, Nd, and Sm, obtained with the respective
elemental abundances listed in Table 7. The dotted lines indicate the
synthesised spectra and the solid lines indicate the observed line
profiles.  Two alternative synthetic spectra for ${\Delta}$[X/Fe] = +0.3
(long dash) and ${\Delta}$[X/Fe] = $-$0.3 (dot-dash) are shown to
demonstrate the sensitivity of the line strength to the abundances.  }
\label{Figure 4 }
\end{figure*}

\begin{figure*}
\epsfig{file=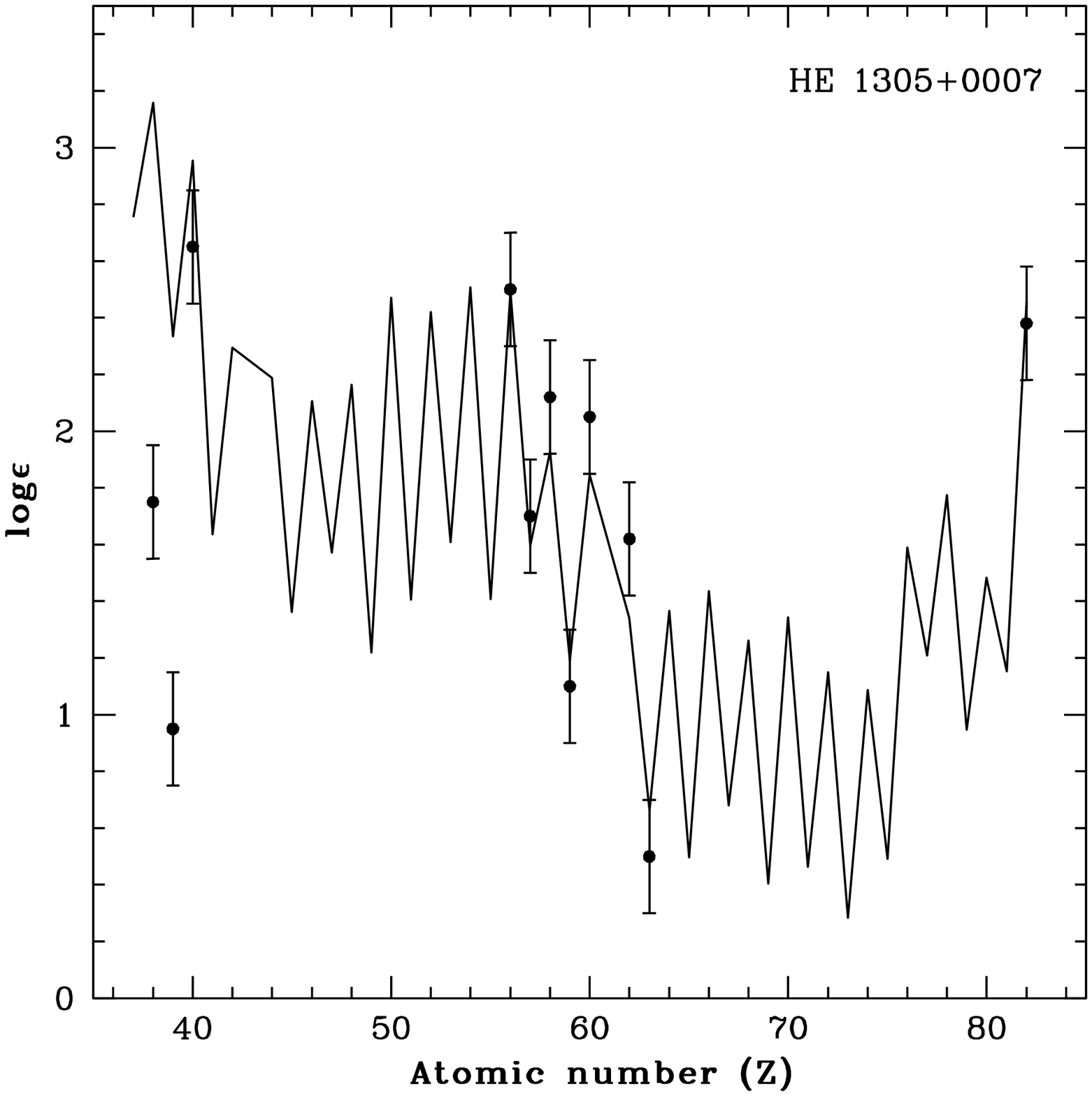,width=14cm,height=8cm,angle=0}
\end{figure*}
\begin{figure*}
\epsfig{file=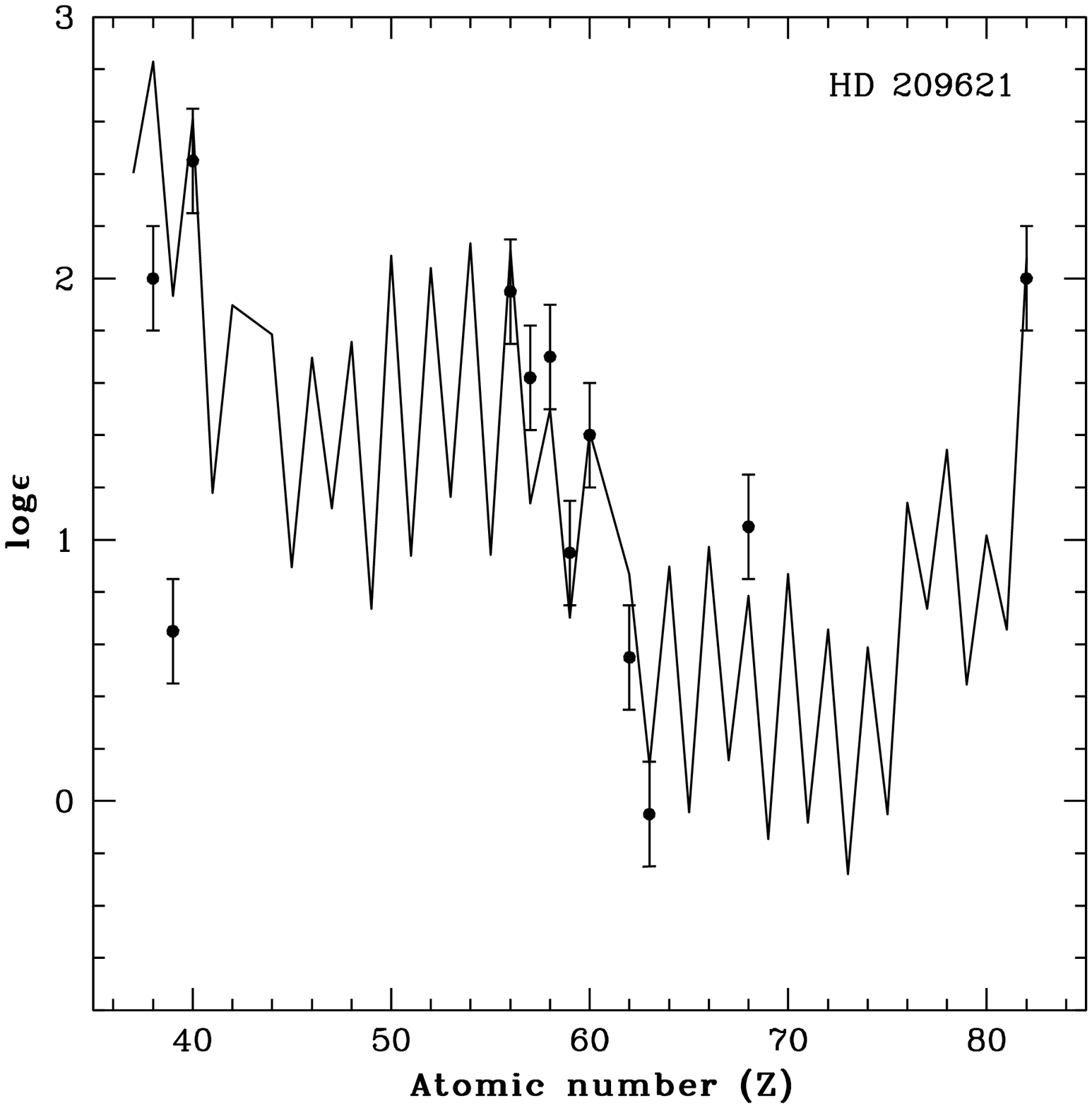,width=14cm,height=8cm,angle=0}
\caption{ Solid curves represent the best fit for the parametric model 
function log${\epsilon}$  = A$_{s}$N$_{si}$ + A$_{r}$ N$_{ri}$, where N$_{si}$
 and N$_{ri}$ represent the abundances 
due to s- and r-process respectively (Arlandini et al. 1999, Stellar model, scaled 
to the metallicity of the star). The best fit coefficients A$_{s}$, A$_{r}$
 and  their respective
 reduced chisquare are listed in Table 8.  The points with errorbars
 indicate the observed abundances in  HE 1305+0007
(upper panel) and HD209621 (lower panel).} 
\label{Figure 5 }
\end{figure*}

\subsection{Error Analysis}

Random errors and systemic errors are the two main sources of errors 
affecting the derived abundances.
 Random errors arise  due to uncertainties 
in line parameters, i.e.  the adopted $gf$ values and the equivalent width
measurements. These  errors  cause line-to-line scatter in derived
abundances for a given species.   Random errors are  minimized by
employing as many usable lines as possible for a given element. In deriving
the Fe abundances we made use of 17 Fe I lines and 4
Fe II lines in HD~209621. The
derived standard deviation ${\sigma}$ is defined by ${\sigma}^{2}$ =
[${\Sigma}(\chi_{i} - {\chi})^{2}/(N-1)$], where $N$ is the number of lines
used. The values of ${\sigma}$ computed from the Fe I lines are ${\pm}$ 0.17
dex in HD~209621. The corresponding value 
calculated for Fe II lines is ${\pm}$ 0.11 dex. 
 The computed errors for Fe I and Fe II are listed in Table 7.

Systemic errors arise from  uncertainties in our adopted 
atmospheric parameters.  The accuracy of the atmospheric
parameters was estimated by computing a set of Fe I lines for pairs of
models with (a) the same gravity and microturbulence velocity but different
temperatures (b) with the same temperature and gravity but different
microturbulence velocities and (c) with the same temperature and
microturbulence velocity but different gravities. Variations in the 
computed equivalent widths from the three cases  are compared
with the accuracy of equivalent width measurements. This comparison 
 allowed us to estimate the uncertainties in the determination of the 
atmospheric parameters. The (conservative) uncertainties in the 
estimated $T_{\rm eff}$ is $\pm 250K$, in surface gravity
 (log\, g) $\pm$ 0.25 dex, and in microturbulence  velocity
($V_{\rm t}$) $\pm$ 0.25 km s$^{-1}$.  Errors in the abundances arising from
the errors in atmospheric parameters are not a simple sum of the errors due
to the individual parameters. These parameters interact with one another
and a change in one may cause a shift in another. The net effect on the
derived mean abundances should be considerably less, because we employ the
principle of consistency wherein the lines with a large range of excitation
potentials, equivalent widths, and different ionization states should lead
us to the same value of abundances.
In Goswami et al. (2006, Table 8), we have shown  the derived differential
abundances of elements with respect to those obtained from the adopted
model  for HE~1305+0007,
by varying $T_{\rm eff}$ by $\pm$ 250
K, log\,g by 0.50 dex, and $V_{\rm t}$ by 0.25 km s$^{-1}$.

Except for Fe abundance, the abundances of all the elements 
listed in Table~7 are derived by
spectrum synthesis calculations where we have visually estimated
the fitting errors. Our estimated fitting errors range between 0.1 dex and
0.3 dex. We adopt these fitting errors as estimates of the random errors
associated with the derived elemental abundances. In Figures 2, 3 and 4
we have shown synthetic spectra for the adopted elemental abundances compared
with the synthetic spectra due to two other possible abundances with
 ${\pm}$ 0.3 dex (see figure captions) differences with
respect to our adopted abundances.

\section{Parametric model analysis of the observed abundances}

To understand the nucleosynthetic origin of the heavy elements it is 
important to know the relative contributions from  the s- and 
r-process  to  their observed abundances. We have examined the 
relative  contributions of these two processes   by comparing the 
observed abundances with predicted s- and r-process contributions 
in the framework of a parametric model for s-process (Howard et al.
 1986). This  model  was also  used by many 
(Aoki et al. 2001, Bo Zhang et al. 2006) to study the s-process element 
abundances in  very  metal-poor stars. We have used the solar system isotopic 
abundances (both s- and r-process) given in   Arlandini et al. (1999). The 
solar system  r- and s- process elemental abundances derived  from the 
isotopic abundances are scaled to  the  metallicity of the star. These 
values (logarithmic with base 10)  are  normalized to the observed Ba 
abundance of the corresponding stars.  

We have used  the following  two parametric model functions \\

 $ N_{i}(Z) = A_{s}N_{s,i} + A_{r}N_{r,i}$  \, \, \, \,(A)\\

 $ N_{i}(Z) = A_{s}N_{s,i} + (1 - A_{s})N_{r,i}$ \, \, \, \,  (B)\\

where Z is the metallicity of the star\\
N$_{s,i}$ is the ith element abundance produced by s-process,\\
N$_{r,i}$ is the ith element abundance produced by r-process. \\

The coefficients A$_{s}$ and  A$_{r}$ obtained from  non-linear least 
square fits
 represent the contributions coming from s- and r-process respectively.
The model fits obtained using model function (A) are shown in figure 5.
The  derived coefficients A$_{s}$,  A$_{r}$ and  the reduced chisquare
($\chi^{2}_{\nu}$) values for HD~209621  are listed in table 8.
These values obtained for HE~1305+0007 is also listed in table 8 for 
comparison. The abundances of neutron-capture elements for HE~1305+0007
are adopted from Goswami et al. (2006).
The 1st peak elements Sr, Y and Zr are not included in the fitting
 as they are not  abundantly produced
as 2nd peak ones by the s-process in metal-poor AGB stars in general. 
The contributions of r- and s-process elements are estimated from the fitting
for Ba-Eu. 

Calculated coefficients  for HE~1305+0007 indicate that contribution from 
the r-process is  higher than  the contribution from s-process; this supports
placing of  HE~1305+0007  in CEMP-r+s group. In case of HD~209621, 
estimated $A_s$  is slightly higher than   A$_{r}$  indicating the
dominance  of s-process in the observed abundances. 
It is noticed in Figure 6, that the abundance pattern of elements 
 56 ${\le}$ Z ${\le}$ 63 agree with the s-process  pattern much better than with the r-process. It is to be noted that,
although  [Ba/Eu] estimate  places HD~209621 in the group of the  CEMP-r+s
stars;  La/Eu and Ce/Eu are high and   that can be explained  by 
s-process models.

{\footnotesize
\begin{table*}
{\bf Table 5: Derived  atmospheric parameters }\\
\begin{tabular}{lcccccc}
\hline
Star Names  &$T_{\rm eff}$ &log\,$g$ & $V_{\rm t}$ km s$^{-1}$ &[Fe I/H]& [Fe II/H] &\\
            &           &         &                    &         &     &    \\
\hline

HD~209621      &  4500   & 2.0   & 2.0  & $-$1.94   & $-$1.92 &   \\
CS~22948-027   &  4750   & 1.5   & 2.0  & $-$2.50   & $-$2.40 &  \\
HE~1305+0007$^{a}$   & 4750    & 2.0   & 2.0  & $-$2.03   & $-$1.99 &  \\
\hline
$^{a}$ from Goswami et al. (2006)\\
\end{tabular}
\end{table*}
}

{\footnotesize
\begin{table*}

{\bf Table 6: Lines used for abundance determination  }\\

\begin{tabular}{ c  c c c c  c }
         &     &     &            &          &   \\

\hline
W$_{lab}$& Z     &  ID  & EP$_{low}$  &log${gf}$ & Remarks  \\
\hline
  5889.951& 11.0 & Na I  &  0.00    & 0.117    &  syn    \\
  5895.932& 11.0 & Na I  &  0.00    & $-$0.184 &  syn    \\
  5172.684& 12.0 & Mg I  &  2.712   & $-$0.402 &   syn   \\
  6155.693& 14.0 & Si I  &  5.619   & $-$1.690  &   syn   \\
  6102.723& 20.0 & Ca I  &  1.879   &  -0.890  &   syn   \\
  4415.559& 21.1 & Sc II &  0.595   &  -0.640  &   syn   \\
  4805.090& 22.1 & Ti II &  2.061   &  -1.100  &   syn  \\
  5247.566& 24.0 & Cr I  &  0.961   &  -1.640  &   syn \\
  5081.107& 28.0 & Ni I  &  3.847   &  +0.300  &   syn \\
  4810.528& 30.0 & Zn I  &  4.078   & -0.137   &   syn, 'as rw'\\
  4607.327& 38.0 & Sr I  &  0.000   &  -0.570  &  syn \\
  5200.413& 39.1 & Y II  &  0.992   &  -0.569  &  syn  \\
  6134.585& 40.0 & Zr I  &  0.000   &  -1.280  &  syn, 'as lw'\\
  6141.727& 56.1 & Ba II &  0.704   &  -0.076  &   syn \\
  5259.380& 57.1 & La II &  0.173   & -1.760   & syn, shallow \\
  5274.229& 58.1 & Ce II &  1.044   &  0.323   &    syn  \\
  5259.728& 59.1 & Pr II &  0.633   &  0.080   &   syn, shallow \\
  5825.857& 60.1 & Nd II &  1.081   & -0.760   &  syn \\
  4791.580& 62.1 & Sm II &  0.104   & -1.44    &   syn  \\
  6437.640& 63.1 & Eu II &  1.319   &  -0.276  &   syn, shallow  \\
  4759.653& 68.0 & Er II &  0.000   & -1.904   &  syn    \\
  4757.542& 74.0 &  W I  &  0.366   & -2.430   &  syn  \\
  4057.807& 82.0 & Pb I  &  1.320   & -0.170   &  syn, 'as'  \\
\hline 
\end{tabular}
\\
``syn" indicates abundances are derived from spectrum synthesis \\
``as lw" indicates asymmetry in the left wing \\
``as rw" indicates asymmetry in the right wing \\
\end{table*}
}

{\footnotesize
\begin{table*}

{\bf {Table 7: Chemical composition of HD~209621 }}

\begin{tabular}{ l c  c c c c c c }
\hline

Element&     $Z$ &Solar$^{a}$&HD~209621      &     &      & $Ref$ & \\ 
      &   &log ${\epsilon}$ &log ${\epsilon}$&[X/H]&[X/Fe]& [X/Fe]  & [X/Ba]\\
\hline
 C           &  6& 8.39&  7.7 & $-$0.69&+1.25   &  +2.29& $-$0.47  \\
 Na I D$_{2}$& 11& 6.17&  4.10& $-$2.07&$-$0.13   & ---   & $-$1.85  \\
 Na I D$_{1}$& 11& 6.17&  4.40& $-$1.77&+0.15   & ---   & $-$1.55  \\
 Mg I        & 12& 7.53&  5.76 &$-$1.77&$+$0.17 & ---   & $-$1.55 \\
 Ca I        & 20& 6.31&  4.45& $-$1.86&$+$0.08 & ---   & $-$1.64 \\
 Sc II       & 21& 3.05&  1.92& $-$1.13&$+$0.79 & ---   & $-$0.91  \\
 Ti II       & 22& 4.9 &  3.70& $-$1.20& +0.72  & ---   & $-$0.98 \\
 Cr I        & 24& 5.64&  3.50& $-$2.14&$-$0.20  & ---   &$-$1.92 \\ 
Fe I        & 26& 7.45&  5.51$\pm 0.17$& $-$1.94& ---& ---  &$-$1.72   \\
Fe II       & 26& 7.45&  5.53$\pm 0.11$& $-$1.92& ---& ---  &$-$1.70  \\
 Ni I        & 28& 6.23& 4.28  &$-$1.95& $-$0.01  & ---   &$-$1.73\\
 Zn I        & 30& 4.60& 2.90& $-$1.70& +0.24  & ----   & $-$1.48  \\
 Sr I        & 38& 2.92& 2.0& $-$0.92& +1.02  & +1.1$\pm 0.3$& $-$0.70  \\
 Y II        & 39& 2.21& 0.65& $-$1.56& +0.36  & ---    & $-$1.34  \\
 Zr  I       & 40& 2.59& 2.45& $-$0.14& +1.80  & ---    & $+$0.08  \\
 Ba II       & 56& 2.17& 1.95& $-$0.22& +1.70  & ---    & ---- \\
 La II       & 57& 1.13& 1.62& $+$0.49& +2.41  & ---    & $+$0.71  \\
 Ce II       & 58& 1.58& 1.70& $+$0.12& +2.04  & +2.8$\pm 0.3$ & $+$0.34  \\
 Pr II       & 59& 0.71& 0.95& $+$0.24& +2.16 & +2.2$\pm 0.7$ & $+$0.46 \\
 Nd II       & 60& 1.45& 1.40& $-$0.05& +1.87  & +2.4$\pm 0.6$ & $+$0.17   \\
 Sm II       & 62& 1.01& 0.55& $-$0.46& +1.46  & ----   & $-$0.24  \\
 Eu II       & 63& 0.52& $-$0.05& $-$0.57& +1.35  & ----   & $-$0.35  \\
 Er II       & 68& 0.93& 1.05& $+$0.12& +2.06  & ---    &  $+$0.34 \\
 W  I        & 74& 1.11& 1.78& $+$0.67& +2.61  & ---    &  $+$0.89 \\
 Pb I        & 82& 2.00& 1.94& $-$0.06& +1.88  & ---    &  $+$0.16 \\
\hline
\end{tabular}
\\
$^{a}$ Asplund, Grevesse \& Sauval   (2005); $Ref$: Vanture (1992c) \\
\end{table*}
}

{\footnotesize
\begin{table*}

{\bf {Table 8 : Coefficients and reduced $\chi^{2}_{\nu}$ for the parametric model
 $ N_{i}(Z) = A_{s}N_{s,i} + A_{r}N_{r,i}$   }}

\begin{tabular}{ l c  c c c c  }
\hline

Objects  &   [Fe/H]   & A$_{s}$  & A$_{r}$  &  ${\chi^{2}_{\nu}}$ \\
\hline
HE 1305+0007 &  -2.0  &  0.469${\pm}$ 0.11 &  0.528${\pm}$0.09 &  1.07 \\
HD 209621    &  -1.94 &  0.568${\pm}$ 0.10 &  0.518${\pm}$0.08 &  1.80 \\
\hline
\end{tabular}
\\
\end{table*}
}

\section{Discussion and concluding remarks }

 We  confirm  HD~209621 to be a  highly carbon-enhanced metal-poor star and 
update earlier  estimates of four s-process elements Y, Ce, Pr and Nd by 
Vanture (1992c). Further, we have obtained error range from 0.1 to 0.3 , 
much lower compared to those of  0.7 and 0.6 dex respectively for Pr and Nd 
by  Vanture (1992c).  New  
abundance estimates for  many other  neutron-capture elements such as Sr, 
Zr, Ba, La, Sm, Eu, Er, W, Pb are presented. The abundance pattern of
 ${\alpha}$ elements are  found to be similar to  those generally seen in 
halo stars of similar metallicity (McWilliam et al. (1995a, b), 
Ryan, Norris, \& Beers (1996), Cayrel et al. 2004). While in Sun 
 [Sr/Ba] = 0.75, HD~209621  shows a much smaller ratio with [Sr/Ba] = -0.68.  

    HD~209621 shows  a larger enhancement of  2nd peak s-process elements 
(Ba, La, Ce, Nd, Sm)  than  1st-peak s-process elements (Sr and Y). 
The enhancement of s-process 
elements  along with the large enhancement of carbon, indicates a 
mass-transfer event in a binary system from a companion AGB star that 
underwent s-process nucleosynthesis during its lifetime. 
The star also shows a large enhancement of Eu  similar to those seen in 
a group of  CEMP-(r+s) stars with [Ba/Eu] ${\ge}$ 0.  
Thus HD~209621 shows characteristics of both CH and 
CEMP-(r+s) stars: in terms of enhancement of heavy s-process elements 
relative to the  lighter s-process elements it shows charecteristics of CH star
 and estimated  [Ba/Eu] = +0.35 places the star in the group of CEMP-(r+s) 
stars. In Figure 7, we have  plotted [Ba/Fe] vs [Eu/Fe] 
for stars  belonging to different classes of CEMP stars.  Different
 classes of CEMP stars are seen to be well separated in the 
 [Ba/Fe]   vs [Eu/Fe] plot. The locations of HD~209621 as well 
as of HE~1305+0007  fall well within the region occupied by 
 (r+s) stars supporting their classification as (r+s) stars.
The chemical homogeneity within the (r+s) group  with a clear 
separation from r- and s-stars is interesting. 

HD~209621 also shows   a large enhancement of the 3rd-peak 
s-process element lead (${\rm [Pb/Fe]} = +1.88$). The 
enhanced Eu  abundance, together with the large abundance of the 3rd-peak 
s-process element Pb indicate a coupling of high s-process and high 
r-process enrichment. 
Enhancement of Pb 
abundance is noticed in a number of  CH as well as  CEMP-(r+s) stars.
Van Eck et al. (2003)  defined stars with [Pb/hs] ${\ge}$ +1.0 
as `lead  stars', where hs includes Ba, La and  Ce.  According  to a 
simplified definition by Jonsell et al. (2006) `lead stars' are those 
with [Pb/Ba] ${\ge}$ 1.0. The latter definition does not require the 
`lead stars'  also to  be (r+s) stars, but requires that they are 
enhanced in Ba and Pb, and much 
more in the latter element. In HD~209621, although we find an enhanced 
abundance of Pb,  [Pb/Ba] is found to be less than +1.0 and hence according
 to the above definitions   HD~209621 does not belong to the group of `lead 
stars'.  Aoki et al. (2001) also found metal-poor stars, LP 625-44 and
 LP 706-7, to be enriched in s-process elements including Pb but cannot be 
considered as lead stars ([Pb/Ce] ${\le}$ 0.4).
In figure 8, we compare the abundances of HD~209621 with the abundances
of  CS~22948-027, CS~22497-034 (Hill et al. 2000,
Barbuy et al. 2005) and HE~1305+0007 (Goswami et al. 2006). These three
stars show large enhancement of Carbon, and   both
r- and s- process elements,  including  Pb. The atmospheric parameters,
metallicities, and heavy-element abundance patterns of these
three objects  are  very similar to HD~209621. Although binarity of
 HE~1305+0007 is not yet established, both  the CS stars and HD~209621
 are known to be long-period binaries (Preston and Sneden 2001, 
Barbuy et al. 2005, McClure and Woodsworth 1990). From the figure it is 
evident that  Pb shows a distinct  trend with metallicity in these objects. 
From the similarity in  the abundances, it is also likely  that these objects 
 form a group   originating  from a similar physical scenario.   

According to the  CEMP stars taxonomy of Beers and Christlieb (2005), (r/s)  
stars are defined by limits on s-elements as [Ba/Fe] ${\ge}$ +1.0 and r-element
 as [Ba/Eu] ${\le}$ 0.5 with [Eu/Fe] ${\ge}$ +1.0. A criteria
 [Ba/Eu] ${\ge}$ 0.0 is set 
for (r+s) stars (Jonsell et al. 2006) which overlaps with the class of `r/s 
stars'. Beers \& Christlieb,  define s-stars as having ${\rm [Ba/Fe]} > +1.0$ 
and  ${\rm [Ba/Eu]} > +0.5$. 
The s-stars are often  C rich (Jonsell et al. 2006, Table 8), while 
the r stars 
are generally less C rich. It is not clear, however, if all (r+s) stars 
are also CH stars, although all 17 (r+s) stars  listed by Jonsell et al.  
have considerably enhanced carbon abundances. However, a relatively high 
fraction of the CH stars seem  to be (r+s) stars (e.g. Aoki et al. 2002). 
According to Abia et al. (2002), CH stars cannot be formed above a threshold 
metallicity, around Z ${\sim}$ 0.4ZM$_{\odot}$; the observed CEMP-(r+s) are 
found to lie 
within a metallicity range -2.0  (HE 1305+0007, Goswami et al. 2006) and
 -3.12 (CS 22183-015, Johnson and Bolte 2002). The majotity of
(r+s) stars listed by Jonsell et al.  are turn-off point stars while 
 CH stars are known to be giants.

In order to understand the origin of neutron-capture elements, the resulting 
abundances of neutron capture elements for HD~209621 are  plotted (figure 6) 
along with the scaled abundance patterns of the solar system material, the 
main s-process component, and  the r-process pattern (Arlandini et al. 1999). 
 The abundance patterns of elements with 
56 ${\le}$ Z ${\le}$ 63  agree with the s -process pattern much 
better than with the r-process pattern indicating  that
the neutron-capture elements in HD~209621 principally originate in the 
s-process. Estimated [Ba/Eu] (= +0.35) for HD~209621 is   
significantly  lower than that  seen in the main s-process component
 ([Ba/Eu] = 1.15, Arlandini et al. 1999). Such low values for [Ba/Eu] were also 
noticed in stars CS~29526-110, CS~22898-027, CS~31062-012, and CS~31062-050;
the ratios  [Ba/Eu], for these objects range from   0.36 to 0.47.  
The abundance patters of elements 
56 ${\le}$ Z ${\le}$ 63 are however in agreement with the s-process 
pattern (Aoki ei al. 2002). 
 The observed low values of [Ba/Eu]  in the four 
CS stars are interpreted  as a result of an s-process that produces 
different abundance ratios from that of the main s-process component 
(Aoki et al. 2002). This interpretation is supported by models 
of  Goriely \& Mowlavi (2000) that
 predicted [Ba/Eu] = 0.4 for yields of metal-deficient AGB stars. 
A similar interpretation also seems applicable to HD~209621; 
 the abundance pattern observed in  
HD~209621   as being produced by s-process although  r-process contamination
 may  have contributed   to  the observed Eu excesses to some extent.
Analysis of the observed abundances of the heavy elements
 with parametric  model function (A) also shows that s-process have  slightly  
higher  contributions  with    $A_s$  = 0.56 
than  r-process with  $A_r$  =  0.51, where   $A_s$ and $A_r$ are the component 
coefficients 
that correspond to contributions  from  the s- and r-process  respectively.
Here,  the contributions from   r- and s-process  are estimated from the 
fitting for Ba - Er.

A widely accepted scenario for the formation of CH stars, known to be
single-line spectroscopic binaries  is the mass transfer from a 
companion AGB star  (McClure 1983, 1984 and McClure and Woodsworth 1990).
Temporal variations of radial velocities observed among the known CH stars
indicate binarity of the objects. 
McClure and Woodsworth  (1990) have  noted the star  HD~209621  to be a 
radial velocity variable with a period of 407.4 days. 
Our estimated radial-velocity  ($-$390 ${\pm}$ 1.5 km s$^{-1}$) also 
 indicates that 
HD~209621 is a high-velocity star and a likely  member of the Galactic halo 
population. Such high velocities are a common feature of CH stars. 
These 
properties  also point  towards the fact, that  the  observed enhancement 
of neutron-capture elements 
in  HD~209621  resulted from a  transfer of material rich in s-process 
elements across a binary system with an AGB star.
The   abundance 
ratios of neutron-capture elements  in  HD~209621 shown by the present study
 underscores the need for detailed studies of a larger sample of CH stars 
in order   to develop a comprehensive scenario for the  production of 
 heavy-elements  by the s-process operating at low metallicity.
Such  studies  are also   likely to provide 
with  clues to the limit on the  metallicity of stars that show double 
enhancement.

\begin{figure*}
\epsfig{file=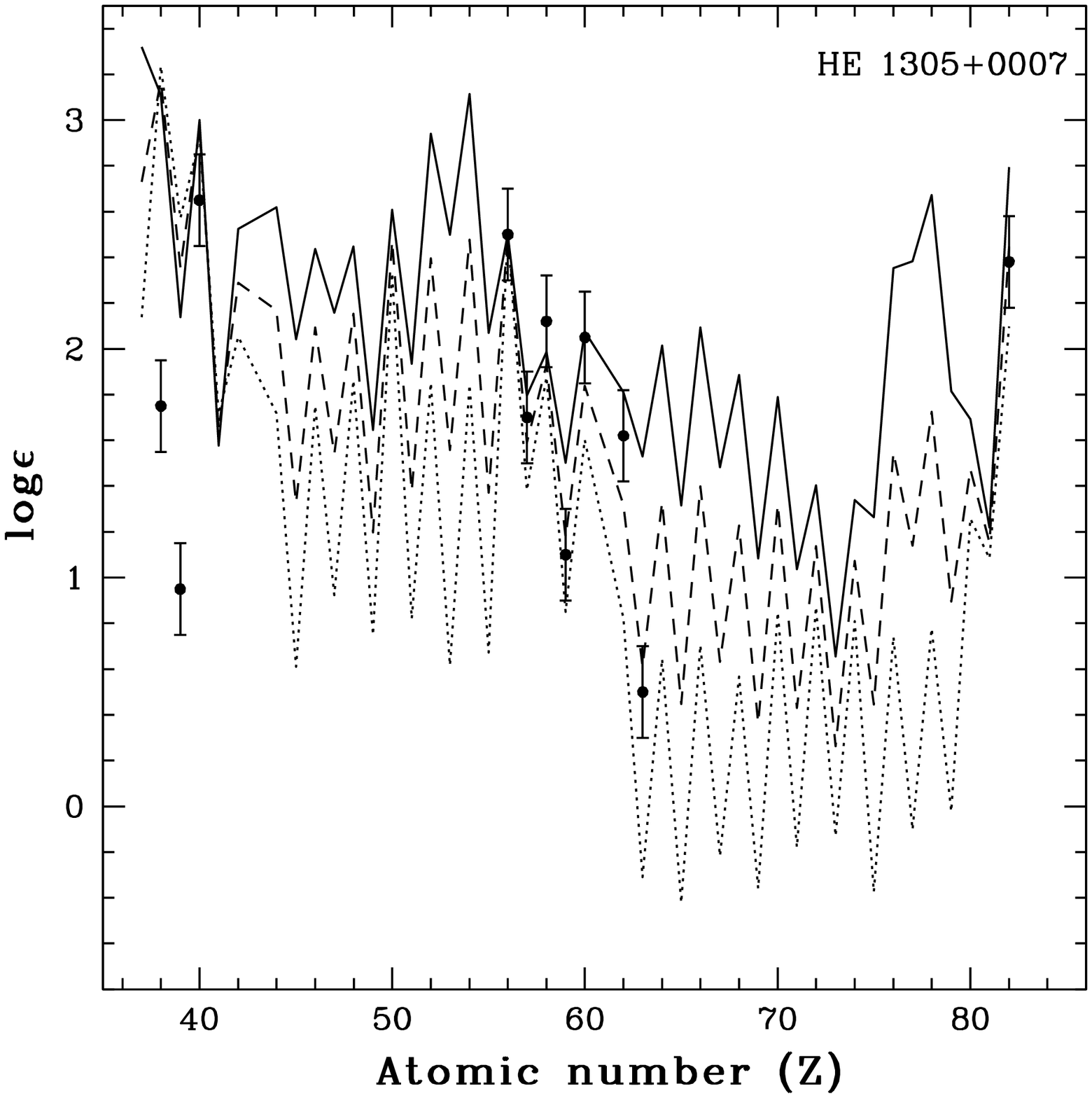,width=14cm,height=8cm,angle=0}
\end{figure*}

\begin{figure*}
\epsfig{file=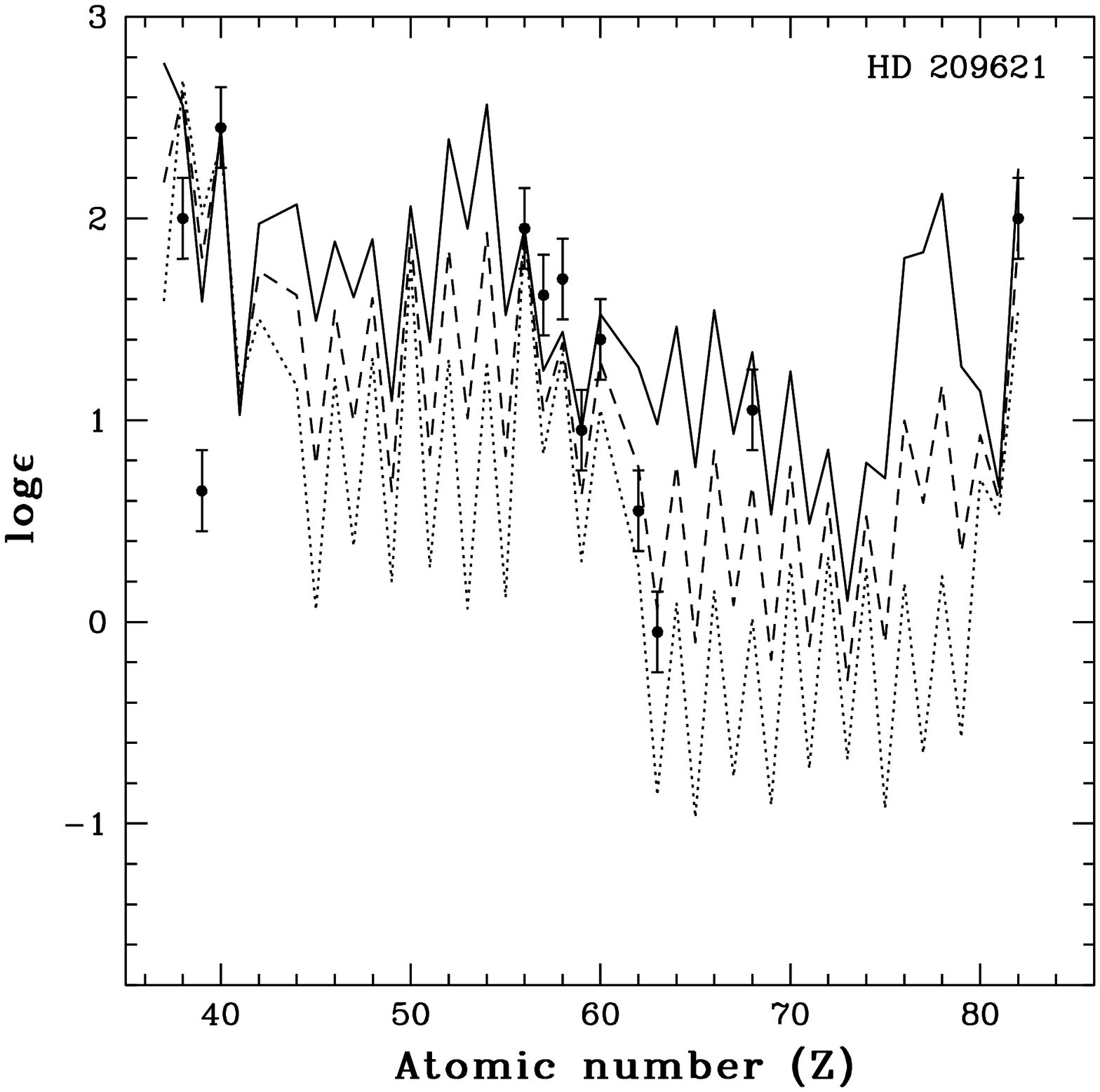,width=14cm,height=8cm,angle=0}
\caption{ 
 Solid line shows abundances due to only r-process, dotted line  
of s-process only and dashed line indicates abundance pattern derived from 
a simple average  abundances coming from  r- and s- process. The  abundances
are scaled to the metallicity of the star and normalized to the observed Ba 
abundances. The points with errorbars indicate the observed abundances in
 HE 1305+0007 (upper panel) and HD 209621 (lower panel).} 
\label{Figure 6 }
\end{figure*}

\begin{figure*}
\epsfxsize= 9truecm
 \epsffile{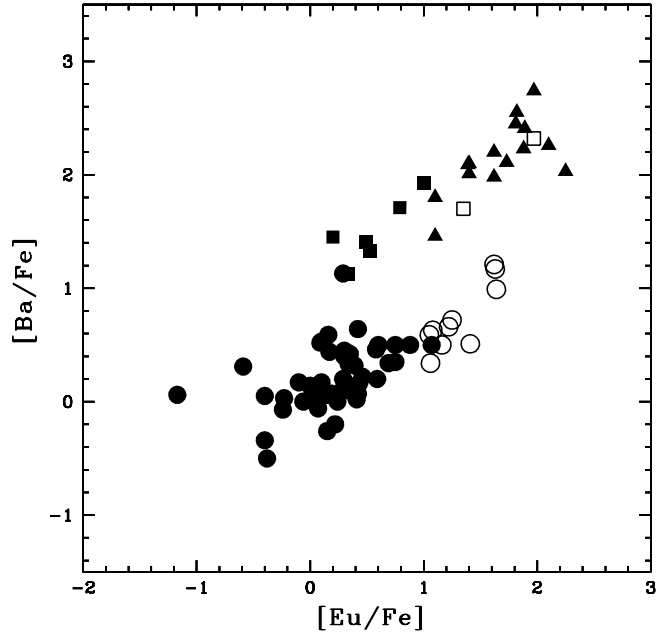}
\caption{ [Ba/Fe] vs [Eu/Fe] plot
 for stars of different classes of CEMP stars. The r+s stars are indicated 
by solid triangles,  r stars by open circles, s stars by solid squares
 and the normal stars that do not show enhancement of either s or r elements 
are shown with solid  circles.
Different classes of CEMP stars are seen to be  well separated; 
 the location of HE~1305+0007 and HD~209621 shown by open  squares are
well within the region occupied by r+s stars. The data points are taken
from the compilation of Jonsell et al. (2006).
  }
\label{Figure 7 }
\end{figure*}

\begin{figure*}
\epsfxsize=9truecm
 \epsffile{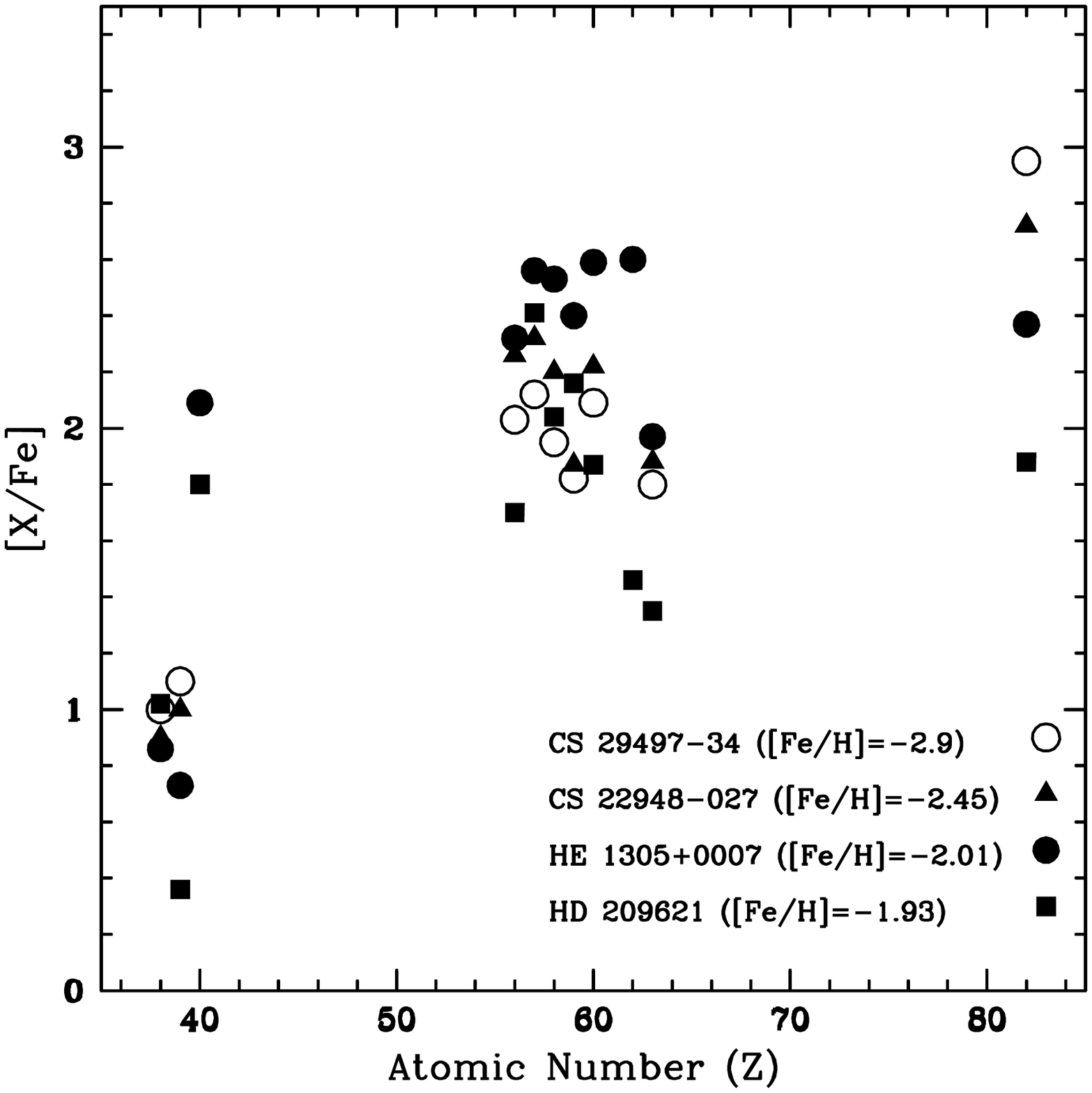}
\caption{A comparison of the distribution of n-capture elements
abundances ([X/Fe]) versus their atomic numbers (Z) in HD~209621 with 
those found in CEMP stars CS~29497-34, CS~22948-027, and HE 1305+0007. The 
later three stars show   double enhancement of  r-  and  s- 
process elements. An interesting feature noticed  is the increase in 
Pb abundances  with  decreasing metallicities  of the stars. 
 }
\label{Figure 8  }
\end{figure*}

{\it Acknowledgements}\\
\noindent
We are grateful to the referee Professor Chris Sneden, for his constructive 
suggestions, which have improved  the readability of the paper considerably.
We would like to thank Drisya Karinkuzhi for figs 5 and 6. This work made 
use of the SIMBAD astronomical database, operated at CDS, Strasbourg, France,
and the NASA ADS, USA.  Funding from the DST Project No. SR/S2/HEP-09/2007 
is gratefully acknowledged. \\
\vskip 1cm

\begin{center}
\bf {\large APPENDIX - A}
\end{center}
{\it Strontium} (Sr)\, The Sr abundance is derived from Sr I line  
at  $4607.327$~{\AA}. Sr II line
at 4077.7 \AA\, is detected as a blended line.
Absorption tip of Sr II line at  4215.52  \AA\, is also identified. This line
is blended with  contributions from the CN molecular band around 4215 \AA\,.
Sr II line at  4161.82 \AA\,  could not be detected in our spectrum. 
Sr I line at 6550.24 \AA\, is   blended  with a Sc II line.

{\it Yttrium} (Y)\, The abundance of Y is derived from  5200.41 \AA\, line.
 Y II line at 4398.1 \AA\, is
asymmetric on the right wing, probably blended with a Nd II line at
4398.013 \AA\, and  could not be used for abundance determination.
Y II line at 4883.69 \AA\, is also asymmetric on the right wing, 
blended with a Sm I line at 4883.777 \AA\,. YII lines  at 5087.43 and 5123.22
\AA\, both are detected, while 5087.43 shows asynmmetry on the right wing, 
5123.22 \AA\,  line is detected as a shallow feature, both the wings   not
reaching the continuum.  YII line at 5205.73 \AA\, is also detected as
a shallow feature.
Y I line at 4982.129 \AA\, is blended with a Mn I line.  The  line 
at 5473.388 \AA\,
is blended with contributions from Mo I and Ce I lines.
Vanture (1992c) estimated Y abundance to be  +1.1 $\pm$ 0.3.  
We note that the  log$gf$ value for 5200.41 \AA\, line used by Vanture is
 $-$0.49. This value is  different from  what we have adopted (-0.569) from 
Sneden et al. (1996).

{\it Zirconium} (Zr)\, The  abundance of Zr is derived using Zr I line
at $6134.57$\,{\AA}. 
Zr II line at 4496.97 \AA\, appears as a blend with Co I
line at 4496.911 \AA\,. Absorption  tip of Zr II line at 4161.21 is easily
detected. Lines  detected at 4208.99, 4258.05 and 4317.32 \AA\, are
heavily contaminated by  contributions from molecular bands.
Zr II line at 4404.75 \AA\, is blended with  an Fe I line and  the 
 4317.321 \AA\,  line 
is  blended with a  Ce II line. None of the Zr II lines are usable
for  abundance determination.

{\it Barium} (Ba)\, We have used the  Ba II line at 6141.727 \AA\,
 to determine the Ba abundance. Although broad, this
line appears as a well defined symmetric line compared to all the other barium
lines detected in the spectrum. log$gf$ value for this line is taken from
Miles and Wiese (1969). 
Ba II line at 4130.65 \AA\,  is found to be
heavily contaminated
by molecular contributions, this line  is not considered for abundance 
determination.
Ba II resonance line at 4554 \AA\, is extremely strong and broad,
 we have excluded this line as well for
abundance analysis. Ba II line at 5853.668 \AA\, shows asymmetry in the right
wing. Ba II line at 6496.897 \AA\, appears as a broad profile much below
the  continuum.  A single
Ba I line that we have detected at 5907.66 \AA\, appears as a 
 broad line and not well defined. We have not used this line for 
abundance determination.

In the case of Ba the hyperfine-splitting (HFS) corrections depend on the
$r/s$ fraction assumed to have contributed to the enrichment of the
star. This element has both odd and even isotopes. The odd isotopes are
mainly produced by the r-process and have a broad HFS, while the even
isotopes are mainly produced by the s-process, and exhibit no HFS. 
It was shown by Sneden et al. (1996) that this issue is important for the
Ba lines at $4554$\,{\AA} and $4934$\,{\AA}, and unimportant for the 
lines at 5854, 6142, and $6497$\,{\AA}.  We have used the red Ba~II line at
$6141.73$\,{\AA}. The HFS
splitting of this line is $\sim$1/5 of the Ba~4554\,{\AA} splitting and
$\sim$1/3 of the thermal line width, and hence  HFS
corrections could be neglected.
 The effect of hyperfine
splitting is smaller than 0.02 dex for the 5853 \AA\, and 6141 \AA\, lines
(McWilliam et al. 1998).
Ba is found to exhibit a marked overabundance of ${\rm [Ba/Fe]} =
+1.70$. 

{\it Lanthanum} (La)\,  The abundance of La is derived from a 
spectrum-synthesis calculation
of the La~II line at $5259.38$\,{\AA}, with atomic data taken from Lawler
et al.  (2001). 
La II line at 4123.23 \AA\,  is detected as a shallow feature. Absorption 
tip of  La II 4238.38 \AA\, 
 line is  detected; this line is  contaminated by molecular contributions. 
 A low 
 absorption profile of 4322.51 \AA\, line is also detected, but not found 
suitable for
abundance determination. The line at 4333.7 \AA\, is contaminated by molecular 
contributions.
La II lines  at 4429.90  and  4558.46 \AA\, both  show
asymmetry in their right wings. La II line at 4574.88 \AA\,  
 also appears as an asymmetric broad profile. Traces of La II lines at
4613.39 and 5123.01 \AA\,  are detected;  low absorption profile of
La II lines at  4662.51 \AA\,    is prominently seen. 
These lines have E$_{low}$ $\sim$ 0 eV. 
Two other zero eV lines  at 4086.709 \AA\, and 5808.313 \AA\, are detected
as shallow profiles with their  wings much below  the continuum. 
6320.43 \AA\, line shows asymmetry on the right wing.

{\it Cerium} (Ce)\,  We have examined seven Ce I and forty five  Ce II
 lines in the spectrun of HD~209621 (Table 9, available on-line); however, they 
are either affected by 
molecular contamination or blended with contributions from other elements. 
The abundance of  Ce is derived from
spectrum-synthesis calculations of the Ce~II line at $5274.23$\,{\AA}. 
Ce exhibits a large
 overabundance of  ${\rm [Ce/Fe]} = +2.04$; this value is  lower than
  Vanture's (+2.8 $\pm$ 0.3).

{\it Praeseodymium}  (Pr)\,  Pr I line at $5996.060$\,{\AA} is detected as an
asymetric line. None of the  Pr II lines in the wavelength region
4100 - 4500 \AA\, (Table 9) were detected in the spectrum of HD~209621. 
Pr II line
at $5188.217$\,{\AA} is blended with a La II line and lines at $5219.045$\,{\AA}
and $5220.108$\,{\AA} appear as  asymetric lines.
The  Pr~II line at $5259.7$\,{\AA} is used to 
derive the abundance of Pr. Pr 
exhibits a large overabundance of ${\rm [Pr/Fe]} = +2.16$. This estimate is
similar to the value of  +2.2 $\pm$ 0.7 derived for Pr by  Vanture (1992c).

{\it  Neodymium } (Nd)\, Nd I line at 6432.680 \AA\, is detected as a 
clean line. Forty two  Nd II lines listed in Table 9  are examined; 
most of them are blended
with contributions from other elements. 
 The abundance  of Nd is  derived  using the  Nd~II line at
5825.85 \AA\,;  Neodymium  shows  overabundance 
 with ${\rm [Nd/Fe]} = +1.87$; this value is also  lower than  Vanture's 
estimate of +2.4 $\pm$ 0.6.  

{\it Samarium }  (Sm)\, Two Sm I and twenty six Sm II lines have been examined
in the spectrum of HD~209621. The abundance of Sm is derived from the 
 Sm~II line at
4791.60 \AA\,. There seems to exist no significant contamination
of this line.  The log$gf$ value is taken from Lawler et al. (2006).
 Sm exhibits an overabundance of ${\rm [Sm/Fe]} = +1.84$. 
 
{\it Europium } (Eu)\,  The main lines of Eu that are
generally used in abundance analysis are the Eu~II lines at 4129.7,
4205.05, 6437.64 and $6645.13$\,{\AA}. The blue Eu~II lines at 4129.7 and
$4205.05$\, {\AA} are severely blended with strong molecular features 
 and could not be used for abundance analysis. The abundance of
Eu  is therefore determined from the red line at
$6437.64$\,{\AA} ( the line at $6645.13$\,{\AA} is severely blended). 
 Spectrum-synthesis
calculations return an abundance of log $\epsilon$(Eu) = 0.04; Eu is 
overabundant with ${\rm [Eu/Fe]} = +1.35$. 

{\it Erbium } (Er)~ Two lines of Er II are identified, 4759.671 \AA\, and 
4820.354 \AA\,. Er abundances listed in Table 7 is estimated from spectrun 
synthesis calculation  of 4759.671 \AA\, line. The line parameters adopted from 
Kurucz atomic line list come from Meggers et al. (1975). Er shows an 
overabundance with [Er/Fe] = +2.06.
Lawler et al. (2008) derived Er abundances in five very metal-poor, 
r-process-rich giant stars,  CS~22892-052, BD~+17~3248, HD~221170, 
HD~115444 and CS~31082-001 and found that the average Er abundance 
for the first four r-process-rich stars is slightly above the scaled
(relative to Eu) solar  value. For CS~31082-001,  Lawler et al. derived 
log ${\epsilon}(Er)$ = -0.30 $\pm$ 0.01 and log ${\epsilon}$(Eu/Er) = $-$0.42
which is  similar  to Eu/Er ratios found for the other four giants.
Estimated Er abundance in CS~29497-030, a star rich in both r- and s-process
material is found to be log ${\epsilon}(Er)$ = -0.57 $\pm$ 0.02 and 
log ${\epsilon}$(Eu/Er) = $-$0.62. The 0.2 dex difference in the ratio 
between this star and the other five stars is interpreted as a result
of  the effects of changing from pure $r$ abundance to a mix of $r+s$.
This value for HD~209621 is ${\sim}$ -1.1.  Lawler et al.  did not use the 
line at 4759.671 {\AA}, this line is  listed  in Kurucz atomic 
line list. The  Er abundances do not  show  noticeable dependence on  
wavelength, log $(gf)$, or on excitation potential (Lawler et al. 2008).

{\it Tungsten } (W)~ Three lines of W I 4757.542 {\AA}, 5793.036 {\AA} 
and 5864.619 {\AA} are identified.  The line used to derive the W 
abundance  is   4757.542 {\AA}.   The abundance derived
is high  with [W/Fe] = 2.61; there is a possible  blend with Cr I line at
4757.578 \AA\,  (with log $g$ and lower excitation potential, -0.920 and 3.55 
respectively) resulting  in an  overestimate of  W abundance.
 
 {\it Lead } (Pb)~  Spectrum-synthesis calculation is also used
to determine the abundance of Pb  using the Pb I line at
$4057.8$\,{\AA}. This line is strongly affected by molecular absorption of
CH. CH lines are included in our spectrum
synthesis calculation.

{}
\end{document}